\documentclass[aps,english,preprint,preprintnumbers,nofootinbib,showpacs]{revtex4-1}
\usepackage{amsfonts,amssymb,amsmath,amsthm,amsxtra}
\usepackage{amsfonts,euscript,mathrsfs}
\usepackage{graphicx}
\usepackage{esint}
\usepackage{hyperref}
\usepackage{latexsym}
\usepackage{epstopdf}
\usepackage{epsf}
\usepackage{epsfig}
\usepackage{bm}%----------bold math
\usepackage{graphicx,color}
\usepackage{dcolumn}%----------Align table columns on decimal point
\usepackage{array}
\usepackage{tabularx}
\usepackage{enumerate}
\usepackage{hhline}
\usepackage{subfigure}
\usepackage{enumerate}
\usepackage{hhline}
\usepackage{babel}

\def\vt{\vartheta}

\makeatletter

\@ifundefined{textcolor}{}
{%
 \definecolor{BLACK}{gray}{0}
 \definecolor{WHITE}{gray}{1}
 \definecolor{RED}{rgb}{1,0,0}
 \definecolor{GREEN}{rgb}{0,1,0}
 \definecolor{BLUE}{rgb}{0,0,1}
 \definecolor{CYAN}{cmyk}{1,0,0,0}
 \definecolor{MAGENTA}{cmyk}{0,1,0,0}
 \definecolor{YELLOW}{cmyk}{0,0,1,0}
 }

\makeatother

\begin{document}

\title{ Scalar-Torsion Mode in a Cosmological Model of the Poincar\'{e} Gauge Theory of Gravity} 

\author{Huan-Hsin Tseng}
\email{d943335@oz.nthu.edu.tw} %
%\affiliation{Department of Physics, National Tsing Hua University, Hsinchu, Taiwan 300}
\author{Chung-Chi Lee}
\email{g9522545@oz.nthu.edu.tw}
%\affiliation{Department of Physics, National Tsing Hua University, Hsinchu, Taiwan 300}
%\affiliation{National Center for Theoretical Sciences, Hsinchu, Taiwan 300}
\author{Chao-Qiang Geng}%$^{2,}$}
\email{geng@phys.nthu.edu.tw}
\affiliation{Department of Physics, National Tsing Hua University, Hsinchu, Taiwan 300}
\affiliation{National Center for Theoretical Sciences, Hsinchu, Taiwan 300}

\date{\today}

\begin{abstract}
We  investigate the scalar-torsion
mode in a cosmological model of the Poincar\'{e} gauge theory of gravity.
We treat the geometric effect of torsion as an effective quantity, which behaves like dark energy,
and study the effective equation of state (EoS) of the model.
 We concentrate on two
cases with the constant curvature solution and positive kinetic
energy, respectively. In the former, we find that the torsion EoS
has different values in the various stages of the universe. In
particular, it behaves like the radiation (matter) EoS of $w_r=1/3$
($w_m=0$)  in the radiation (matter) dominant epoch,
while in the late time the torsion density is supportive for the accelerating
universe. In the latter, our numerical analysis shows that in general
the EoS has an asymptotic behavior in the high redshift regime,
while it could cross the phantom divide line in the low redshift regime.

\end{abstract}

%PACS:   98.80.Jk, 04.50.?h, 98.80.Es

\maketitle

\section{Introduction}

Recent cosmological observations~\cite{obs1,obs11,obs12,obs13,obs14}
have demonstrated that our universe is undergoing the phase of an accelerating expansion.
Although general relativity (GR) developed in the last century has been successful in many ways
of explaining various experimental results in gravity,  the nature of the accelerating universe now
rises as a small cloud shrouding it. We thereby look for a more
general theory that comprises GR yet able to solve the
accelerating problem referred to as dark energy~\cite{DE}.
By virtue of the local gauge principle, one leads to incorporate Poincar\'{e}
group as gauge group of a principal bundle, such that the local Lorentz symmetry of
the spacetime is preserved~\cite{Hehl:1976kj}.
An attempt is to release torsion from a connection, rather than
the Levi-Civita connection in the standard GR, which also acts as a
dynamical field like metric tensor. Such spacetime
is usually called Riemann-Cartan manifold.
The gauge theory based on this manifold,
 is known as Poincar\'{e} gauge theory (PGT)~\cite{Obukhov:1987tz,Hehl:1979xk,Hayashi:1981mm,Hehl:1994ue}.

It has been investigated in~\cite{Hayashi:1981mm,Sezgin:1979zf}  that there are six
modes by the decomposition of the connection according to torsion
tensor in the linearlized theory, classified as $0^{\pm}$, $1^{\pm}$ and
$2^{\pm}$ in terms of spins and parities.
Among them,
%the $0^-$ mode~\cite{Kopczynski}, acting as a pseudoscalar, is considered to be the only dynamical field of the axial
%torsion. While matter composed of fermions, coupling to its intrinsic
%spin, is the source to interact with this mode.
%Since the effect of the axial torsion is taken to be small in
%general,  the $0^-$ mode is less studied in the literature.
%On the other hand,
the $0^+$ mode~\cite{Kopczynski}, also called the {\it scalar-torsion} mode,
does not directly interact with any known fundamental source~\cite{Shie:2008ms}.
Along with the property induced by the nonlinear
equation set, this $0^+$ mode is our main concern.
In~\cite{Shie:2008ms}, Shie, Nester and Yo (SNY) have examined models with
the spin $0^+$ mode in PGT to achieve the late time accelerating expansion of the universe.
In other words,  the geometric effect of torsion is treated as an effective dark energy.
In particular, they have presented  two cases with the solutions of  a constant curvature
and positive kinetic energy, respectively. The first case is
an extremely simple solution existing inside the system of
differential equations formed by the spin $0^+$ mode, which
provides a modeling for the late time de Sitter universe for
dark energy.
Note that this simple solution violates the positivity argument~\cite{Shie:2008ms}.
The second one is referred to as the normal case,
which conforms with the regular positivity condition, but it gives rise
to no obvious analytic solution.
Torsion cosmology related to the scalar-torsion mode has been also explored
in~\cite{Chen:2009at,Li:2009zzc,Li:2009gj,
Ho:2011qn,Ho:2011xf,Ao:2010mg,Baekler:2010fr,Ao:2011kc,Xi:2011uz}.
In this work, we concentrate on these two cases and
 present numerical solutions of the late-time
acceleration behavior corresponding to the equation of state (EoS),
%%%%%%%%%%%
% {\color{red}
% playing an important role in cosmological evolution and
% }
%%%%%%%%%%%%%%%%%
defined by $w=p/\rho$, where $\rho$ and $p$ are the energy density
and pressure of the relevant component of the universe, respectively.

This paper is organized as follows:  In Sec.~II, we review
the scalar-torsion of the spin $0^+$ mode in PGT and give equations of motion for cosmology.
In Sec.~III, we show our numerical results on
the cosmological evolutions for the scalar-torsion mode.
We present our conclusions in Sec.~IV.

\section{Scalar-torsion Mode in Poincar\'{e} Gauge Theory}

\subsection{Lagrangian for the scalar-torsion mode}

PGT of gravity starts with a Lagrangian 4-form on
$U_4$-spacetime:
\begin{equation}\label{E:Lagrangian}
\mathcal{L}(g,\vt,\Gamma) = \mathcal{L}_G + \mathcal{L}_{M}\,,
\end{equation}
where $\{\vt^i\}$ is a set of the orthonormal dual basis,
$\Gamma_j{}^i$ is the connection 1-form with respect to $\{\vt^i\}$,
$\mathcal{L}_M$ is the matter Lagrangian, and $\mathcal{L}_G$ is the
gravitational Lagrangian that can be made up by certain
combinations. In~\cite{Shie:2008ms}, SNY studied the spin
$0^+$ mode, given by~\cite{Shie:2008ms,Hehl:2012pi}
\begin{equation}\label{E:SNY}
\mathcal{L}_{G} =  \frac{a_0}{2} R \eta + \frac{b}{24} R^2 \eta +
\frac{a_1}{8} \left( {}^{(1)} T^i \wedge \star{}^{(1)} T_i \right)\,,
\end{equation}
where $\star$ is the Hodge dual map, $\eta$ is the volume 4-form of
the space-time and ${}^{(J)} T^i$ with $J=1,2,3$ are the irreducible
pieces of the torsion 2-form $T^i= d\vt^i + \Gamma_j{}^i \wedge
\vt^j$, defined by~\cite{Hehl:1994ue}
\[
{}^{(1)} T^i= T^i - {}^{(2)}T^i - {}^{(3)}T^i, \qquad {}^{(2)}T^i=
\frac{1}{3} \vt^i \wedge (i_{e_j} T^j), \qquad {}^{(3)}T^i =
\frac{1}{3} \star\left( \vt^i\wedge \star (T^j \wedge \vt_j) \right)\,.
\]
The coefficients of $\mathcal{L}_G$ in (\ref{E:SNY}) are  constrained
 by the positivity argument~\cite{Shie:2008ms} such
that
\begin{equation}\label{E:condition}
a_1 > 0 ,\qquad b > 0.
\end{equation}
The independent variation of (\ref{E:Lagrangian}) with respect to
$(g_{ij},\vt^i,\Gamma_j{}^i)$ yields~\cite{Trautman:2006fp}
\begin{eqnarray}
\delta \mathcal{L}_G &=& \frac{1}{2} K^{ij} \, \delta g_{ij} + E_i \wedge \delta
\vt^i + E_i{}^j \wedge \delta \Gamma_j{}^i + \mbox{ an exact form},\\
\delta \mathcal{L}_M &=& \frac{1}{2} T^{ij} \, \delta g_{ij} + t_i
\wedge \delta \vt^i + s_i{}^j \wedge \delta \Gamma_j{}^i + \mbox{ an
exact form},
\end{eqnarray}
where the gauge field momenta are given by
\begin{equation}\label{E:momenta}
K^{ij}= 2 \, \frac{\delta \mathcal{L}_G}{\delta g_{ij}}, \qquad E_i =
\frac{\delta \mathcal{L}_G}{\delta \vt^i}, \qquad E_i{}^j= \frac{\delta
\mathcal{L}_G}{\delta \Gamma_j{}^i},
\end{equation}
and the source terms are defined by
\begin{equation}\label{E:momenta}
T^{ij}= 2 \, \frac{\delta L_M}{\delta g_{ij}}, \qquad t_i =
\frac{\delta L_M}{\delta \vt^i}, \qquad s_i{}^j= \frac{\delta
L_M}{\delta \Gamma_j{}^i},
\end{equation}
corresponding to the symmetric energy-momentum tensor-valued 4-form,
asymmetric vector-valued 3-form usually called canonical
energy-momentum tensor, and tensor-valued 3-form known as canonical
spin angular momentum tensor, respectively. One can also write the
decompositions into the basis of $\Omega(M)$~\cite{Trautman:2006fp,Hehl:2012pi}:
\begin{equation}\label{E:canonical EM tensor}
t_i = \mathscr{T}_{ik} \, \eta^k, \qquad s_{ij} = S_{ijk} \, \eta^k,
\end{equation}
where $\eta^k := \star \, \vt^k$. The equations of motion are given
symbolically as
\begin{equation}\label{E:EOM}
E_i = -t_i, \qquad E_{ij} = - s_{ij}.
\end{equation}

\subsection{Equations of motion for cosmology}

We shall describe our universe with the FLRW cosmology, which is
homogeneous and isotropic with the metric
\begin{equation}\label{E:FRW}
ds^2 = -dt^2 + a^2(t) \left( \frac{dr^2}{1-kr^2} + r^2 d\Omega^2\right),
\end{equation}
where $k$ is the constant curvature.
% of the time slices $S(t)$, $M= I_{a(t)} \times S$. However, f
For simplicity, we shall
only consider the flat universe with $k=0$.

 For (\ref{E:SNY}) of  the SNY model  in the FRLW cosmology with
no spin source $(S_{ijk} \equiv 0)$, the main field equation
(\ref{E:EOM}) leads to~\cite{Shie:2008ms}
\begin{eqnarray}
\label{E:main eq1}
\dot{H} = \frac{\mu}{6 a_1} R + \frac{1}{6a_1} \mathscr{T} -2H^2,\\
\label{E:main eq2}
\dot{\Phi}(t) = \frac{a_0}{2a_1}R + \frac{\mathscr{T}}{2 a_1} - 3 H
\Phi + \frac{1}{3} \Phi^2,\\
\label{E:main eq3} \dot{R} = -\frac{2}{3} \left( R + \frac{6\mu}{b}
\right) \Phi,
\end{eqnarray}
where $\mu = a_1 + a_0$, $H= \dot{a}(t)/a(t)$, and
$\Phi(t)= T_t$, which is the time component of the torsion trace, defined by
$T_i=T_{ij}{}^j$.
%,  constituting the coefficients of ${}^{(2)}T^i$.
Here, $R$ in (\ref{E:main eq1})-(\ref{E:main eq3})
denotes the affine curvature with respect
to the curvature 2-form $\Omega_i{}^j$, given by
\begin{equation}
\Omega_i{}^j = d\Gamma_i{}^j + \Gamma_k{}^j \wedge \Gamma_i{}^k =
\frac{1}{2} R^j{}_{ikl} \, \vt^k \wedge \vt^l.
\end{equation}
Hence, one obtains the relation
\begin{equation}
R = \bar{R} + 2 \frac{\partial T^j}{\partial x^j} - \frac{2}{3} T_kT^k\,,
\end{equation}
where $\bar{R} = 6(\dot{H} + 2H^2)$ represents  the curvature of the
Levi-Civita connection induced by (\ref{E:FRW}). The energy-momentum
tensor $\mathscr{T}_{ij}$ is defined as (\ref{E:canonical EM
tensor}) and $\mathscr{T}$ stands for  the trace
$\mathscr{T}_i{}^i$. Explicitly, one has
\begin{equation}\label{E:rho_T}
\begin{aligned}
\mathscr{T}_{tt} &= \rho_{M} = \frac{b}{18} \left( R + \frac{6\mu}{b}
\right) \left( 3H - \Phi \right)^2 - \frac{b}{24}R^2 - 3a_1 H^2,\\
\mathscr{T} &= 3p_{M}- \rho_{M} \,.
\end{aligned}
\end{equation}
with the subscript $M$ representing the ordinary matter including
both dust and radiation.

%In \cite{Shie:2008ms}, they treat 
To see the geometric effect of torsion,
% as an effective quantity out of GR. Therefore one can succeed the insight descendent from GR, as seen from 
we can write down the Friedmann equations as
\begin{eqnarray}
H^2 &=& \frac{\rho_c}{3a_0}, \qquad  \quad \quad \quad \rho_c =\rho_M + \rho_T,\nonumber \\
 \dot{H} &=& -\frac{\rho_c + p_{tot}}{2 a_0}, \qquad p_{tot} = p_M + p_T,
\label{E:FriedmannEq}
\end{eqnarray}
with $a_0=\left( 8 \pi G \right)^{-1}$ in GR, where $\rho_c$ and $p_{tot}$
denote the critical energy density and total pressure of the
universe, while $\rho_T$ and $p_T$ correspond to the energy density and
pressure of some effective field, respectively. 
By comparing the equation of
motion of the scalar-torsion mode in PGT~(\ref{E:rho_T}) to the
Friedmann equations~(\ref{E:FriedmannEq}), one obtains
\begin{eqnarray}\label{E:rho,p_T}
\rho_T &=&  3\mu H^2 - \frac{b}{18} \left( R + \frac{6\mu}{b} \right)
(3H - \Phi )^2 + \frac{b}{24} R^2,\nonumber\\
p_T &=& \frac{1}{3} \left( \mu( R - \bar{R} ) + \rho_T \right),
\end{eqnarray}
which will be regarded as
%It is clear that one may regard the
%contributions from the torsion as 
the effective 
%energy density and pressure. So that the 
torsion dark energy density
and pressure, respectively.
% are then natural to define as the following,}
%where
%\begin{eqnarray}
%H^2 &=& \frac{\rho_c}{3a_0}, \qquad  \quad \quad \quad \rho_c =\rho_M + \rho_T,\nonumber \\
% \dot{H} &=& -\frac{\rho_c + p_{tot}}{2 a_0}, \qquad p_{tot} = p_M + p_T,
%\end{eqnarray}
%with the subscript $M$ representing the ordinary matter including
%both dust and radiation.
%%From (\ref{E:rho,p_T}), we can define the EoS of torsion dark energy as
%%\begin{equation}\label{E:w_T}
%%w_T= p_T/\rho_T.
%%\end{equation}
%$\bar{R}_{\mu \nu}$ and $\bar{R}$
 From~(\ref{E:FriedmannEq}), we get the continuity
equation,
% the definitions above and the fact that
%}
% we deduce
\begin{equation}\label{E:conservation_rho_c}
\dot{\rho}_c + 3H \left( \rho_c + p_{tot} \right) =0,
\end{equation}
%which indicates that the continuity equation holds for the whole matter,
%{\color{red}
which can also be derived by applying the identity 
\[
\bar{\nabla}_j
\bar{G}^{ij} = \bar{\nabla}_j \left( \bar{R}^{ij} - \frac{1}{2}
\bar{R} g^{ij} \right)= \bar{\nabla}_j \left(
\mathscr{T}^{ij}+\mathscr{T}_T^{ij} \right) =0, 
\]
where
$\bar{\nabla}$ is the covariant derivative with respect to the
Levi-Civita connection and $\mathscr{T}_{T \  j}^{\ i}=diag\left(
-\rho_T, p_T, p_T, p_T \right)$ is the effective energy-momentum
tensor of torsion dark energy. 

In addition, from (\ref{E:main eq1}) - (\ref{E:main eq3}), one can
check that
%With another important fact to note is that
the continuity equation for the torsion field is also valid, $i.e.$
\begin{equation}
\label{E:conservation_rho_T}
\dot{\rho}_T + 3 H \left( \rho_T+ p_T \right) =0.
\end{equation}
%as one can check by using ~(\ref{E:main eq1}) - (\ref{E:main eq3}).
%Therefore (\ref{E:conservation_rho_c}) and
%(\ref{E:conservation_rho_T}) tells us
%Clearly,
Consequently, we obtain the continuity equation for the ordinary
matter to be
\begin{equation}\label{E:conservation_rho_M}
\dot{\rho}_M + 3 H \left( \rho_M+ p_M \right) =0.
%, \qquad \mbox{($w_M = 0 \;or\; \frac{1}{3}$)}
\end{equation}
%serves independently for itself.
By assuming no coupling between radiation and dust, the matter densities of radiation ($w_r=1/3$) and dust ($w_m = 0$)
in scalar-torsion cosmology share the same evolution behaviors  as GR, $i.e.$ $\rho_r \propto a^{-4}$ and
$\rho_m \propto a^{-3}$, respectively.
In order to investigate the cosmological evolution, it is natural to define the total EoS by~\cite{DE}
\begin{eqnarray}
w_{tot} = -1 - \frac{2 \dot{H}}{3 H^2} =\frac{p_{tot}}{\rho_c},
\end{eqnarray}
which leads to 
\begin{eqnarray}
\label{E:w_T}
 w_{tot} =\Omega_M w_M + \Omega_T w_T,
\end{eqnarray}
where $\Omega_\alpha=\rho_\alpha/\rho_c$ and $w_\alpha = p_\alpha/\rho_\alpha$ with $\alpha=M,T$,
representing the energy density ratios and EoSs of matter and torsion, respectively.   
%, so that with $\Omega_T = \rho_T / \rho_c$ it deduces a canonical identification that
%\begin{equation}\label{E:w_T}
%w_T= p_T/\rho_T,
%\end{equation}
Note that the EoS in (\ref{E:w_T}), which is commonly used in the literature, e.g.~\cite{DE},  
can be examined by cosmological observations in~\cite{obs1,obs11,obs12,obs13,obs14}.
In particular, it can be used to distinguish the modified gravity theories from $\Lambda$CDM~\cite{DE}.
%such as type-Ia
%supernovae (SNIa), baryon acoustic oscillation (BAO) and cosmic
%microwave background radiation (CMB).} 
%On the other hand, this definition~(\ref{E:w_T}) is commonly used in the literature,
%e.g.~\cite{DE}. }
%In order to investigate
%% the evolution behavior of
%the torsion density, we {\color{red}commonly} follow \cite{DE} to define the ratio of the pressure to the energy density as the effective dynamical EoS of
%the torsion dark energy,
%\begin{equation}\label{E:w_T}
%w_T= p_T/\rho_T.
%\end{equation}
Consequently, the evolution of the torsion dark energy can be described in
terms of solely $w_T$ by
\begin{eqnarray}
\rho_T(z) = \rho_T^{(0)} \exp \left\{ 3 \int_0^z dz^{\prime} \frac{1+w_T(z^{\prime})}{1+z^{\prime}} \right\}.
\end{eqnarray}
In the following section, we shall focus on
%discussing
this important quantity.

\section{Numerical Results of Torsion Cosmology}
The evolution of torsion cosmology is determined by (\ref{E:main
eq1}) - (\ref{E:main eq3}). In general, one needs to solve the
dynamics of $R$, $\Phi$ and $H$ by the system of ordinary
differential equations. However, one easily sees that in
(\ref{E:main eq3}) there exists a special case, in which the constant
scalar affine curvature $R=-6\mu/b$ is a possible solution~\cite{Shie:2008ms}.
Recall that in order to conform with the positive
kinetic energy argument, the condition (\ref{E:condition}) is
needed. However, since the special case yields the
negative curvature  $R=-6\mu/b <0$ with a negative matter density
$\rho<0$,
the condition of $a_1 < -a_0 <0$
is required~\cite{Shie:2008ms}.

In this section, we concentrate on the EoS of the scalar-torsion mode
in both special and normal cases.
  We also present the cosmological evolution of
the density ratio, defined by $\Omega=\rho/\rho_c$, from a high
redshift to the current stage.

\subsection{Special Case: $R= const.$}\label{sec:specialcase}
In this special case, we take the assumptions of  $a_1, \mu<0$ and $a_0>0$
in~\cite{Shie:2008ms}. The evolution equations (\ref{E:main eq1}) -
(\ref{E:main eq3}) reduce to
\begin{eqnarray}
\label{eq:sc01}
\rho_M &=& - 3 a_1 H^2 - \frac{3}{2}\frac{\mu^2}{b}, \\
\label{eq:sc02}
\rho_T &=& \frac{3}{2}\frac{\mu^2}{b} + 3 \mu H^2, \\
\label{eq:sc03} \dot{H} &=& -\left( 1+w_M \right)
\left(\frac{3}{4}\frac{\mu^2}{a_1 b}+\frac{3}{2}H^2\right).
\end{eqnarray}

For the numerical calculation, we rescale the
parameters as follows:
\begin{eqnarray}
\label{eq:screscaling}
 m^2& =&\rho_m^{(0)}/3 a_0\,, \quad
 %\nonumber \\
 \tilde{a}_0 = a_0/m^2 b, \quad \tilde{a}_1 = -a_1/m^2 b, \nonumber \\
\tilde{t}&= & m \cdot t, \quad \quad \tilde{\mu}=
\tilde{a}_1-\tilde{a}_0, \quad
 \tilde{H}^2 =H^2/m^2, \quad \tilde{R}=R/m^2,
\end{eqnarray}
where $\rho_m^{(0)}$ is the matter density at $z=0$ and the scalar
affine curvature is a constant $\tilde{R}= 6 \tilde{\mu} >0$. From
(\ref{eq:sc01}), (\ref{eq:sc02}) and (\ref{eq:sc03}),
% divided by $m^2$,
 we obtain the following dimensionless equations,
\begin{eqnarray}
\label{eq:scHm2}
&&\tilde{H}^2= \frac{\tilde{a}_0}{\tilde{a}_1}\left( a^{-3}+\chi a^{-4} \right) + \frac{\tilde{\mu}^2}{2\tilde{a}_1}, \\
\label{eq:scrhoT}
&&\frac{\rho_T}{\rho_m^{(0)}}= \frac{ \tilde{\mu}^2}{2 \tilde{a}_0}- \frac{ \tilde{\mu}}{\tilde{a}_0} \tilde{H}^2, \\
\label{eq:scdotH}
&&\tilde{H}{\tilde{H}}^{\prime}= \left( 1+w_M \right) \left(\frac{3}{4} \frac{\tilde{\mu}^2}{\tilde{a}_1}- \frac{3}{2} \tilde{H}^2\right)\,,
\end{eqnarray}
where the prime $``\prime''$ stands for  $d/d\ln a$ and
$\chi=\rho_r^{(0)}/\rho_m^{(0)}$. Using (\ref{E:conservation_rho_T}),
%The EoS $w_T$
(\ref{E:w_T})
% is found from the continuity equation ${\rho}^{\prime}_{T} + 3 (1+w_T)\rho_T = 0$
 and (\ref{eq:scrhoT}), we find that
\begin{eqnarray}
\label{eq:scTEOS}
w_T = -1 - \frac{\dot{\rho}_{T}}{3H \rho_T}= -1 -\frac{4}{3} \frac{\dot{\tilde{H}}}{2\tilde{H}^2-  \tilde{\mu}}.
\end{eqnarray}

\begin{center}
\begin{figure}[tbp]
\begin{tabular}{ll}
\begin{minipage}{80mm}
\begin{center}
\unitlength=1mm
\resizebox{!}{6.5cm}{\includegraphics{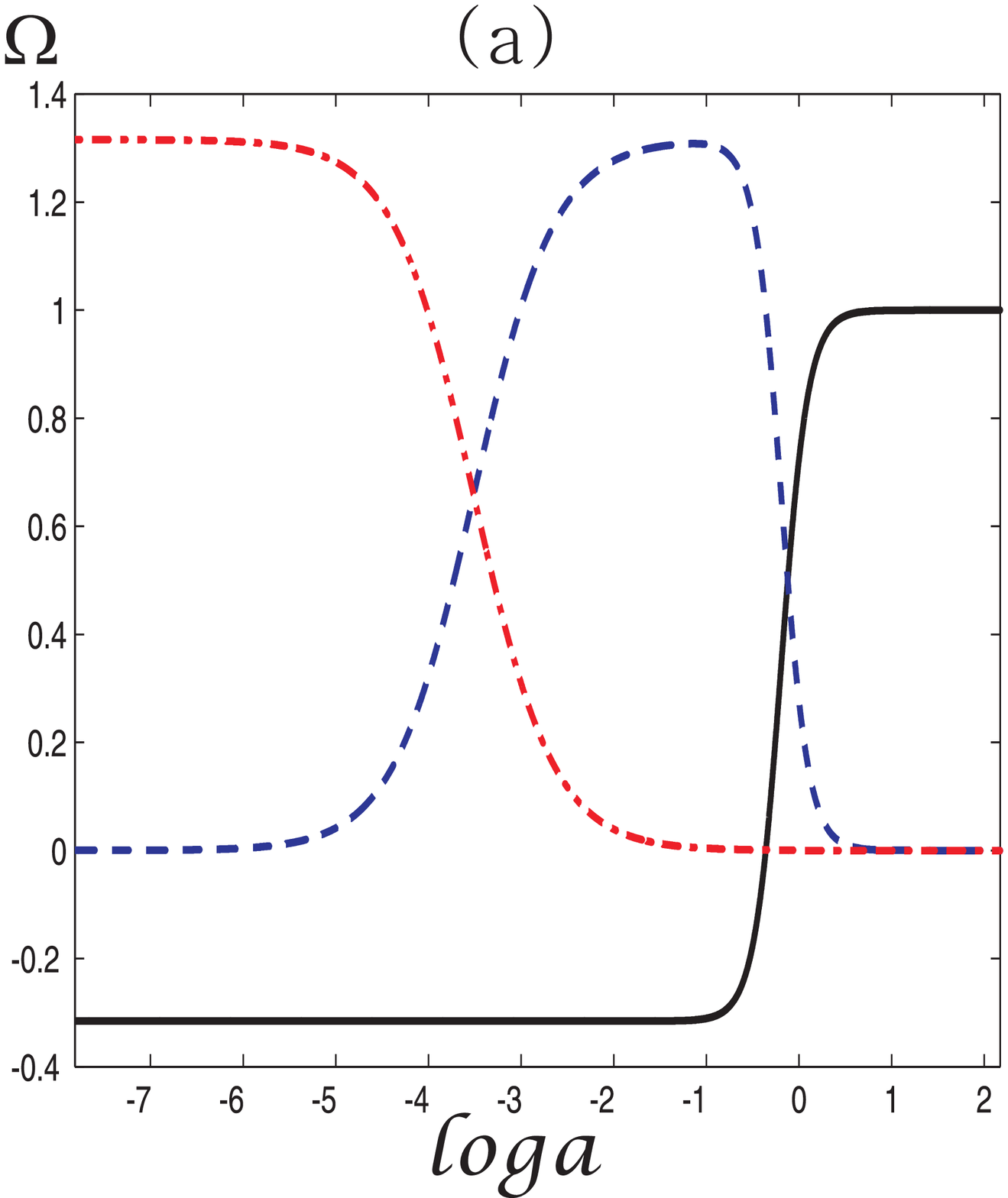}}
\end{center}
\end{minipage}
&
\begin{minipage}{80mm}
\begin{center}
\unitlength=1mm
\resizebox{!}{6.5cm}{\includegraphics{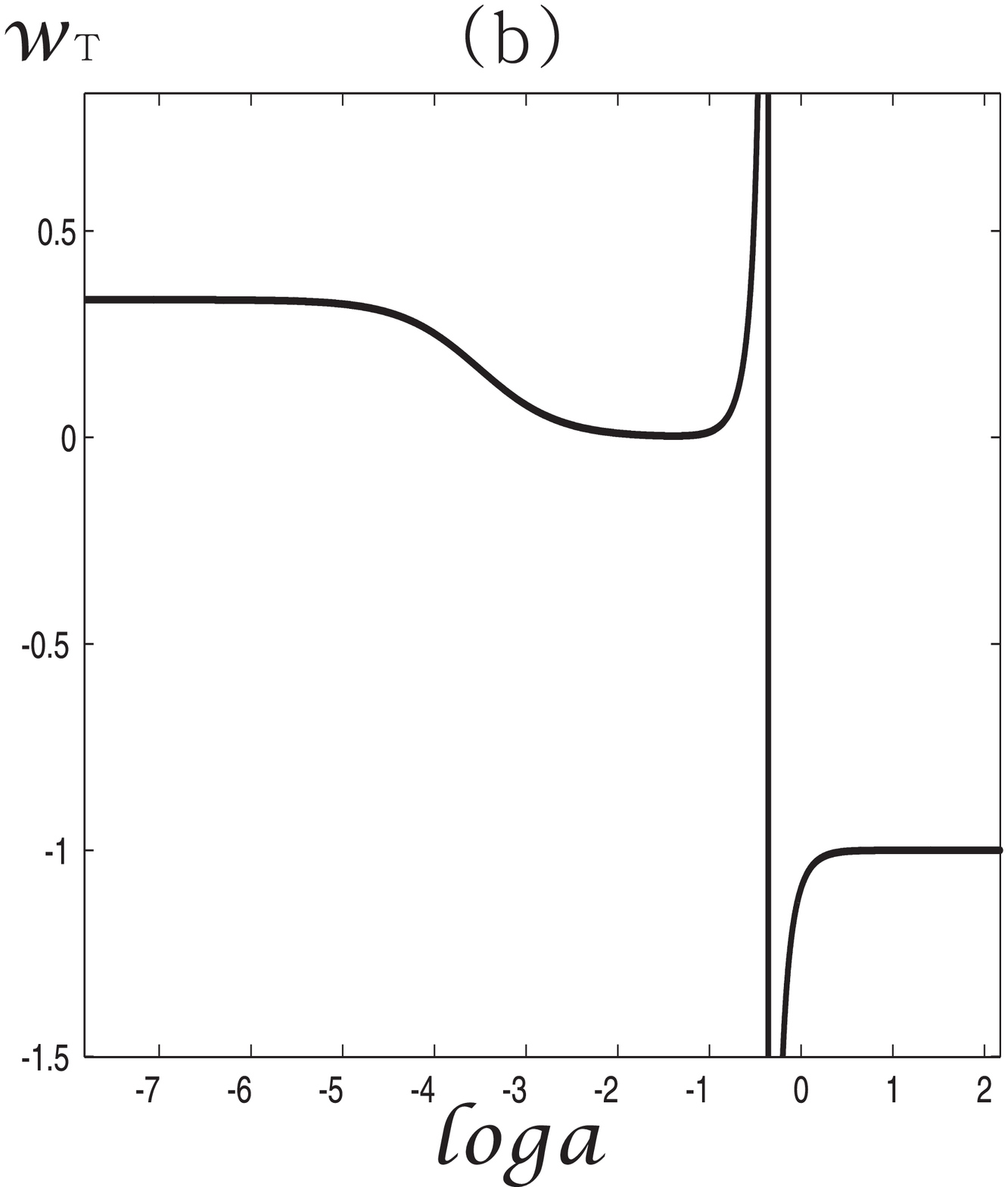}}
\end{center}
\end{minipage}\\[5mm]
\end{tabular}
\caption{
Evolutions of (a) the energy density ratio $\Omega$  and (b) the torsion EoS $w_T$
with $\Omega_m^{(0)}=27.5\%$,
where the solid (black), dashed (blue), and dotted-dashed (red) lines stand
for torsion, matter and radiation, respectively. }
\label{fg:1}
\end{figure}
\end{center}

  From (\ref{eq:scHm2})-(\ref{eq:scTEOS}), it is easy to see
  that the evolution of $\rho_T$
is automatically determined without solving any differential
equation for given values of
$\tilde{a}_0$ and $\tilde{a}_1$.
 The numerical results of this special case are shown in
Fig~\ref{fg:1}, where we have chosen $\tilde{a}_0=76$,
$\tilde{a}_1=100$ and $\chi=3.07 \times 10^{-4}$ in corresponding to
$\Omega_m^{(0)}=\tilde{H}^{-2}_{z=0}\simeq 27.5\%$. In
Fig.~\ref{fg:1}a, we plot the energy density ratios of torsion, matter and
radiation, $\Omega_T$, $\Omega_m$ and $\Omega_r$, respectively.
Notice
that $\rho_T$ depends on the parameters $\tilde{a}_0$ and $\tilde{a}_1$,
and there exists a late-time de-Sitter solution when $\tilde{H}^2=
\tilde{\mu}^2 / 2 \tilde{a}_1$. In the high redshift regime, in
which $\tilde{H}^2 \gg \tilde{\mu}, \tilde{\mu}^2/ \tilde{a}_1$,
we observe that the torsion density ratio $\Omega_T$ is a constant
which can also be estimated from (\ref{eq:sc01}) and
(\ref{eq:sc02}), namely
\begin{eqnarray}
\label{eq:scMoT} \frac{\rho_M}{\rho_T} = \frac{3 \tilde{a}_1
\tilde{H}^2 - 3\tilde{\mu}^2/2}{3\tilde{\mu}^2/2 - 3
\tilde{\mu} \tilde{H}^2} \simeq -\frac{\tilde{a}_1}{\tilde{\mu}},
\end{eqnarray}
which manifests itself a negative constant. In Fig.~\ref{fg:1}b, we
show that the torsion EoS $w_T$
%and find that
  acts  as matter $w_m=0$ and radiation $w_r=1/3$
in the matter-dominant  ($\rho_m \gg \rho_r$) and
%, while it behaves like radiation $w_r=1/3$ in the
radiation-dominant  ($\rho_r \gg \rho_m$) stages, respectively,
which are interesting asymptotic behaviors.
We also observe that
in the low redshift regime of $\log \, a \simeq 0$, $w_T$ is smaller
than unity, indicating the existence of a late-time acceleration
epoch.

\subsection{Normal Case}\label{sec:normalcase}

The normal case here denotes the positive definiteness of both
kinetic energy and matter density, i.e, the parameters $a_0$, $a_1$
and $b$ are subject to the condition (\ref{E:condition}). It is also
convenient to rescale the parameters such that
\begin{eqnarray}
\label{eq:norescaling}
\tilde{a}_0 &=& a_0/m^2b, \quad \tilde{a}_1 =a_1/m^2b, \quad
\tilde{t} =t \cdot m, \quad  \tilde{\mu} =\tilde{a}_0 + \tilde{a}_1, \nonumber \\
\tilde{H}^2& =&H^2/m^2, \quad \tilde{\Phi} =\Phi/m, \quad
\tilde{R}=R/m^2,
\end{eqnarray}
where $m^2=\rho_m^{(0)}/3 a_0$. Using the above rescaling
parameters, (\ref{E:main eq1}) - (\ref{E:main eq3}) and (\ref{E:rho_T}) are then rewritten
as
\begin{eqnarray}
\label{eq:no1}
&&\tilde{H}{\tilde{H}}^{\prime}= \frac{\tilde{\mu}}{6\tilde{a}_1} \tilde{R} - \frac{\tilde{a}_0}{2\tilde{a}_1} a^{-3} -2\tilde{H}^2, \\
\label{eq:no2}
&&\tilde{H}{\tilde{\Phi}}^{\prime}=\frac{\tilde{a}_0}{2\tilde{a}_1}\left( \tilde{R} -3a^{-3} \right) -3\tilde{H}\tilde{\Phi} + \frac{1}{3}\tilde{\Phi}^2,\\
\label{eq:no3}
&&\tilde{H}{\tilde{R}}^{\prime}= - \frac{2}{3} \left( \tilde{R} + 6 \tilde{\mu} \right) \tilde{\Phi},\\
\label{eq:no4}
&&\frac{1}{18}\left( \tilde{R} + 6 \tilde{\mu} \right) \left( 3 \tilde{H} - \tilde{\Phi} \right) - \frac{\tilde{R}^2}{24} -3\tilde{a}_1 \tilde{H}^2 = 3 \tilde{a}_0 \left( a^{-3}+\chi a^{-4} \right),
\end{eqnarray}
where
we have used
$\overset{em}{T}=3P_M-\rho_M= - \rho_m = - 3 a_0
m^2 a^{-3}$ due to
$w_r = p_r / \rho_r = 1/3$ and $w_m = p_m / \rho_m = 0$.
%, which leads to}.
 From $(\ref{E:w_T})$ and (\ref{eq:no1})-(\ref{eq:no4}),
we have
\begin{eqnarray}
\label{eq:no_w} w_T = \frac{1}{3} \,  \frac{\tilde{\mu} \left(
\tilde{R}- \bar{R}/m^2 \right)}{3\tilde{\mu} \tilde{H}^2 - \left(
\tilde{R} + 6 \tilde{\mu} \right) \left( 3 \tilde{H} - \tilde{\Phi}
\right)^2 / 18 + \tilde{R}^2/24} +\frac{1}{3}.
\end{eqnarray}

\begin{center}
\begin{figure}[tbp]
\begin{tabular}{ll}
\begin{minipage}{80mm}
\begin{center}
\unitlength=1mm
\resizebox{!}{6.5cm}{\includegraphics{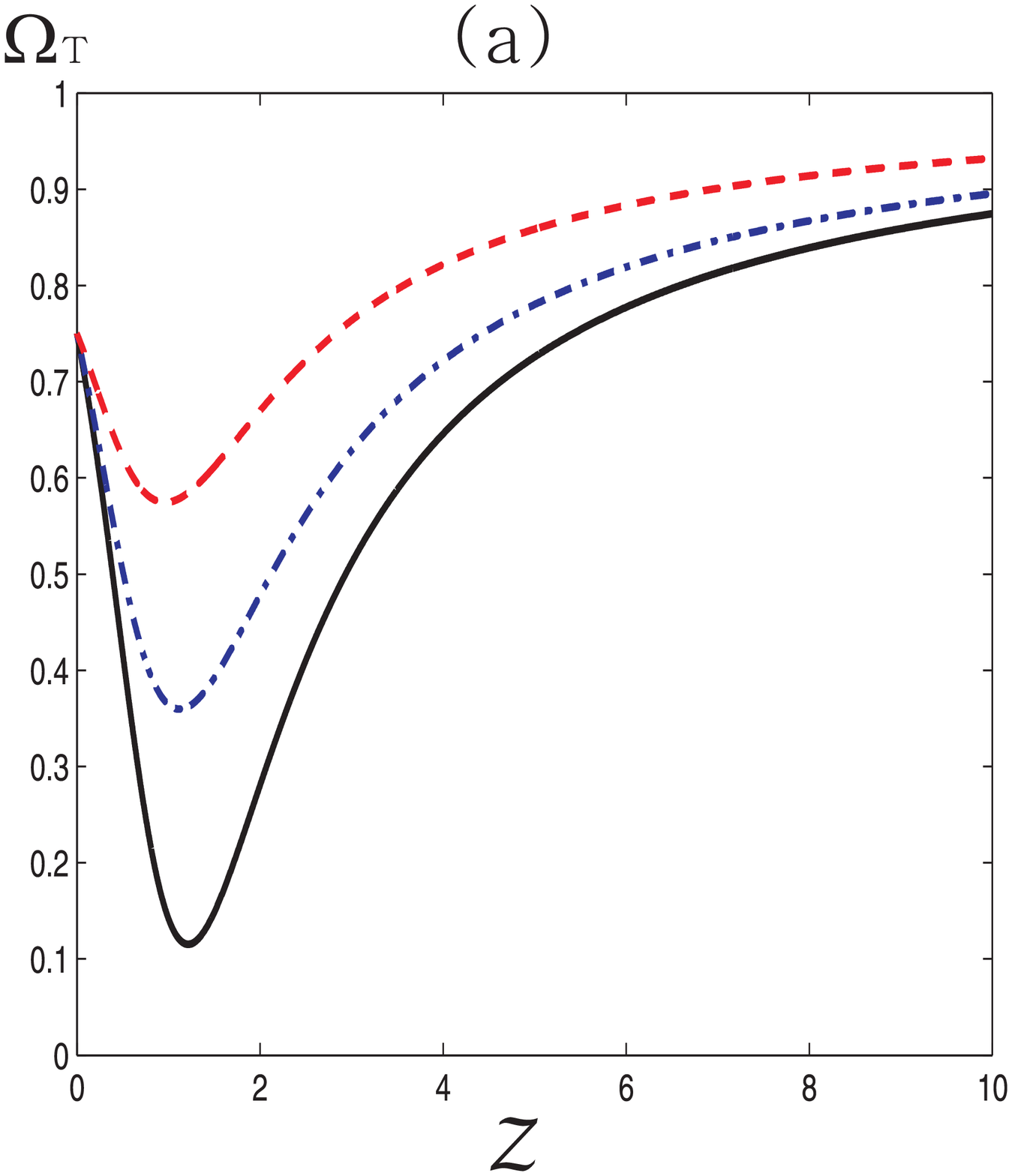}}
\end{center}
\end{minipage}
&
\begin{minipage}{80mm}
\begin{center}
\unitlength=1mm
\resizebox{!}{6.5cm}{\includegraphics{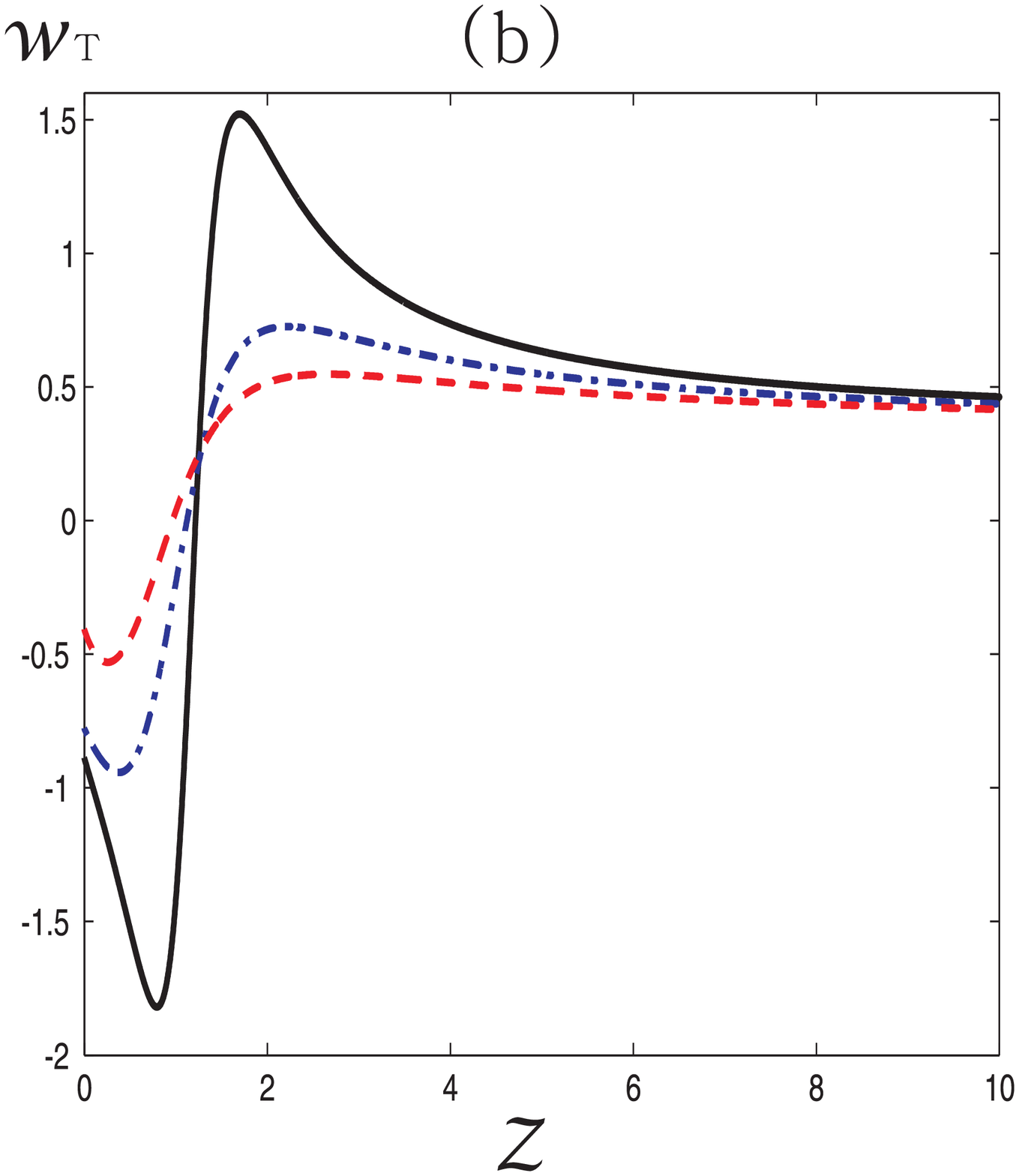}}
\end{center}
\end{minipage}\\[5mm]
\end{tabular}
\caption{Evolutions of (a) the energy density ratio $\Omega_T$ and (b) the
torsion EoS $w_T$ in the universe as functions of the redshift $z$
with $\Omega_m^{(0)}=25\%$ and $\chi=3.07 \times 10^{-4}$, where the
solid, dotted-dashed and dashed lines
correspond to
 $ (\tilde{a}_0, \tilde{a}_1, \tilde{R}_0, \tilde{H}_0) = (2, 1, 14, 2), (2, 1, 13, 2), (3, 1, 8, 2)$, respectively.}
\label{fg:2}
\end{figure}
\end{center}

To perform the numerical computations, we need to specify two
parameters: $\tilde{a}_0$ and $\tilde{a}_1$, along with two initial
conditions: $\tilde{R}$ and $\tilde{H}$. Thus, the initial condition
for $\tilde{\Phi}$ is automatically determined by (\ref{eq:no4}).
The numerical results are shown in Fig.~\ref{fg:2}, where the
initial conditions at $z=0$ are set as $(\tilde{a}_0,
\tilde{a}_1, \tilde{R}_0, \tilde{H}_0)
 = (2, 1, 14, 2), (2, 1, 13, 2), (3, 1, 8, 2)$ for
solid, dot-dashed, and dashed lines, respectively. Note
that  $\chi=3.07 \times 10^{-4}$ originates from the
WMAP-5 data, and $\tilde{H}=2$ corresponds to $\Omega_m^{(0)}
=\tilde{H}^{-2}_0 = 0.25$.

In  Fig.~\ref{fg:2}a, we  show the evolution of the density ratio,
$\Omega_T = \rho_T / \rho_c$, as a function of the redshift $z$. The
figure demonstrates that the torsion density $\rho_T$ dominates the
universe in the high redshift regime ($z \gg 1$) with the general
parameter and initial condition selection, while the
matter-dominated regime is reached only within a very short time
interval. In Fig.~\ref{fg:2}b, we show that $w_T$ has an
asymptotic behavior at the high redshift regime, $i.e.$ $ w_{z \gg
0} \rightarrow 1/3$. Moreover, in the low redshift regime, it may
even have a phantom crossing behavior, $i.e.$, the torsion EoS could
cross the phantom divide line of $w_T=-1$. As a result, the
scalar-torsion mode is able to account for the late-time
accelerating universe.
%, but is hard to obtain the similar result as
%the standard $\Lambda$CDM model.
Finally, we remark that the studies in Refs.~\cite{Shie:2008ms,Ao:2010mg} only indicated
an oscillating behavior without the asymptotic one above for the torsion density.

\section{Conclusions}
\label{sec:conclusion} 
We have studied the torsion EoS of the two
cases of the scalar-torsion mode in PGT of gravity, which are
suitable for explaining the late-time accelerating universe but each
of them possesses a quite different cosmological behavior in the
high redshift regime. For the first case,
%investigate the special case,
which violates the positive kinetic energy and has a
constant affine curvature $R$, the torsion EoS  has  asymptotic
behaviors: $w_T=1/3$ in the radiation-dominated stage,
$w_T=0$ in the matter-dominated stage,
 and finally a late-time de-Sitter
solution corresponding to $w_T=-1$. The torsion density ratio of
 $\Omega_T$ in the high redshift regime is a ``negative''
constant.
For the second one, which has  the positive kinetic energy,
under the general selection of parameters and
initial conditions, the torsion EoS still shows an asymptotic behavior, $w=1/3$,
in the high redshift regime, while it could cross the phantom divide line in the
low redshift regime. The most confusing phenomenon in this spin
$0^+$ scalar-torsion cosmology is that the universe is dominated by torsion
in the high redshift regime even though there exists a narrow window
for the matter-dominated epoch.
% Even though there  might be dominated by matter, the torsion density dominated the universe in the final (high redshift) region.

\begin{acknowledgments}
 We are grateful to Professor H.~J.~Yo and Professor J.~M.~Nester
 for the inspiring and helpful discussions.
This work was partially supported by National Center of Theoretical
Science, National Tsing Hua University and  National Science Council (NSC-98-2112-M-007-008-MY3) of R.O.C.
\end{acknowledgments}

\end{document}